\documentclass[prd,aps,nofootinbib,floatfix,11pt]{revtex4}

\usepackage{amsmath,graphicx,epsfig,amssymb,dsfont,mathtools}
\usepackage[usenames]{color}
\usepackage{ulem} 
\usepackage{bigstrut}
\usepackage{slashed}
\usepackage{multirow}
\usepackage{subfigure}

\allowdisplaybreaks

\begin{document}
	
\title{The semi-leptonic form factors of $\Lambda_{b}\to\Lambda_{c}$ and $\Xi_{b}\to\Xi_{c}$ in QCD sum rules}
	
\author{
Zhen-Xing Zhao$^{1}$~\footnote{Email:zhaozx19@imu.edu.cn},
Run-Hui Li$^{1}$~\footnote{Email:lirh@imu.edu.cn},
Yue-Long Shen$^{2}$~\footnote{Email:shenylmeteor@ouc.edu.cn},
Yu-Ji Shi$^{3}$~\footnote{Email:shiyuji92@126.com}
Yan-Sheng Yang$^{1}$,
}
	
\affiliation{
$^{1}$ School of Physical Science and Technology, \\
Inner Mongolia University, Hohhot 010021, China \\
$^{2}$ College of Information Science and Engineering, \\
Ocean University of China, Qingdao 266100, P.R. China \\
$^{3}$ Helmholtz-Institut f\"ur Strahlen- und Kernphysik and Bethe Center
for Theoretical Physics, Universit\"at Bonn, 53115 Bonn, Germany
}
		
\begin{abstract}
In this work, the full leading order results of the form factors for $\Xi_{b}\to\Xi_{c}$ and $\Lambda_{b}\to\Lambda_{c}$ are obtained in QCD sum
rules. Contributions from up to dim-5
have been considered. For completeness, we also study the two-point
correlation function to obtain the pole residues of $\Xi_{Q}$ and
$\Lambda_{Q}$, and higher accuracy is achieved. For the three-point correlation function, since stable Borel regions can not be found, about
$20\%$ uncertainties are introduced for the form factors
of $\Xi_{b}\to\Xi_{c}$ and $\Lambda_{b}\to\Lambda_{c}$.
Our results for the form factors are consistent with those of the
Lattice QCD within errors.
\end{abstract}

\maketitle

\section{Introduction}

The study of semi-leptonic decay $\Lambda_{b}\to\Lambda_{c}\ell \nu$ is of great
phenomenological significance as it provides an ideal place to constrain the CKM matrix element
$V_{cb}$. Furthermore, this process can also play an important role to test the lepton universality.
The measured branching ratio is given by \cite{Tanabashi:2018oca}
\begin{equation}
{\cal B}(\Lambda_{b}^{0}\to\Lambda_{c}^{+}e^{-}\bar{\nu}_{e})=(6.2_{-1.3}^{+1.4})\times10^{-2}.
\end{equation}
To extract $V_{cb}$ or test the lepton universality, one must have the knowledge of $\Lambda_b \to \Lambda_c$ transition form factors, which are defined as
\begin{align}
&\langle \Lambda_{c}(p_{2},s_{2})|\bar{c}\gamma_{\mu}(1-\gamma_{5})b|{\Lambda_b}(p_{1},s_{1})
\rangle \nonumber \\ =&\bar{u}(p_{2},s_{2})\left[\gamma_{\mu}f_{1}(q^{2})+i\sigma_{\mu\nu}\frac{q^{\nu}}{M_{1}}f_{2}(q^{2})
+\frac{q_{\mu}}{M_{1}}f_{3}(q^{2})\right]u(p_{1},s_{1})\nonumber \\ -&\bar{u}(p_{2},s_{2})\left[\gamma_{\mu}g_{1}(q^{2})+i\sigma_{\mu\nu}\frac{q^{\nu}}{M_{1}}g_{2}(q^{2})
+\frac{q_{\mu}}{M_{1}}g_{3}(q^{2})\right]\gamma_{5}u(p_{1},s_{1}).\label{eq:parametrization_standard}
\end{align}
In the heavy quark limit, the form factors $f_1$ and $g_1$ reduce to one unique Isgur-Wise function $\zeta(w)$ , where $w=v\cdot v'$, and $f_2=f_3=g_2=g_3=0$.  At zero recoil, we have $\zeta(1)=1$. The heavy quark effective theory (HQET) provides a systemical framework to study the power corrections  to the predictions in the heavy quark limit.

When the  recoil energy is  small, Lattice QCD simulation works well and there already exist predictions of the $\Lambda_b \to \Lambda_c$ form factors \cite{Detmold:2015aaa}, while one has to employ phenomenological models to extrapolate the result to the whole momentum region. It makes great sense to evaluate the form factors in the large recoil region as  the model dependence will be effectively reduced. Some works based on various quark models have been done \cite{Ivanov:1999pz,Albertus:2004wj,Faustov:2016pal,Zhao:2018zcb,Zhu:2018jet,Ke:2019smy,Becirevic:2020nmb,Thakkar:2020vpv}, while they are highly model dependent. Perturbative QCD approach (PQCD) is adopted in \cite{Shih:1999yh}, but a relative small  branching fraction of about 2\% is obtained for $\Lambda_b \to \Lambda_c \ell \bar{\nu}$. In \cite{Guo:2005qa}, HQET and PQCD are adopted at small recoil and large recoil region respectively, and the diquark picture is used for $u,d$ quarks. 

The QCD sum rules method  is a time-honored QCD-based approach to deal with hadronic parameters. It reveals a direct connection between hadron phenomenology and QCD vacuum structure via a few universal parameters such as quark condensates and gluon condensates. The method has been successfully applied to various problems relevant to the hadron structures. The three-point QCD sum rules have been widely used in the study on the transition form factors. For the heavy-to-light form factors such as $B \to \pi$ form factors, the light-cone sum rules is more appropriate because the light-cone dominance of the correlation functions is proved at the large recoil region. While for the heavy-to-heavy case, the three-point QCD sum rules are applicable if the virtuality of the momentum of the interpolating current is sufficient large (LCSRs are also applicable at appropriate virtuality region). In \cite{Shi:2019hbf}, we derived the form factors of doubly heavy baryons to singly heavy baryons using QCD sum rules for the first time, but so far our results are hard to be tested for the lack of experimental data. For the singly heavy baryon decays more data are accumulated which can help to check the theoretical predictions.  In this work we will calculate the $\Lambda_{b}\to\Lambda_{c}$ and $\Xi_{b}\to\Xi_{c}$ transition form factors with three-point QCD sum rules so that the validity of the three-point sum rules can be checked.

Most studies on the heavy-to-heavy or heavy-to-light transition form factors are based on HQET (for the $\Lambda_b \to \Lambda_c$ form factors, some of them can be found in \cite{Grozin:1992mk,Dai:1996xv,Wang:2003it,Huang:2005mea}, while \cite{Wang:2009yma} is based on the light-cone QCD sum rules), thus the power suppressed contributions are neglected. In this paper, we will employ the heavy quark field in full QCD.  In this respect two studies have already been performed \cite{MarquesdeCarvalho:1999bqs,Azizi:2018axf}, however, there are large discrepancy between these two works. The form factors obtained in \cite{Azizi:2018axf} seems not reasonable because the form factors at the small recoil do not meet the predictions of HQET. While for \cite{MarquesdeCarvalho:1999bqs}, there are some places to be improved, one is the two Borel parameters are not taken as free parameters, and the other is the following predictions for the form factors defined in Eq. (\ref{eq:parametrization_standard})
\begin{equation}
f_{1}=g_{1},\quad f_{2}=f_{3}=g_{2}=g_{3}=0,
\end{equation}
is too rough. For the latter, the authors have only adopted the coefficients of the Dirac structures with the highest dimension to extract the vector and axial-vector form factors. In fact, at the next-to-leading power (NLP) of $1/m_{Q}$ in HQET, $f_{2}$ and $g_{3}$ are fairly large rather than zero. Therefore, a more careful study on the $\Lambda_b \to \Lambda_c$ transition is required. In addition, when \cite{MarquesdeCarvalho:1999bqs} was done, there was no mature Lattice QCD calculation available. In this work, we will make close comparisons with the predictions of the Lattice QCD.

\begin{figure}[!]
\includegraphics[width=0.3\columnwidth]{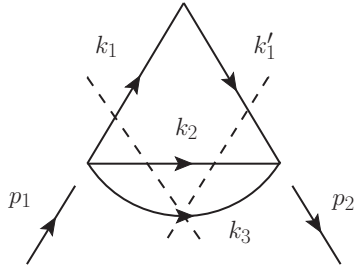}
\caption{Cutting rules for the spectral density of the double dispersion relation in Eq. (\ref{eq:3p_correlator_QCD}).}
\label{fig:ff_cutting}
\end{figure}

In our method there are two points that need to be emphasized \cite{Shi:2019hbf}. The first one is  to obtain the spectral densities of double dispersion relations using cutting rules. As can be seen in Fig. \ref{fig:ff_cutting},   all the propagators perpendicular to $p_{1}$ are to be cut if we intend to take discontinuity with respect to $p_{1}^{2}$, and the same is true for the case of $p_{2}$. This is readily justified using numerical integration for the situation of scalar quarks, and can be evidently proved with the approach provided in \cite{Veltman:1994wz}. The other one is to deal with the superfluous Dirac structures by taking into account the contributions from the negative parity baryons. Because we are interested in the process of $1/2^{+}\to1/2^{+}$, once the other 3 processes including $1/2^{-}$ baryons are considered, all the coefficients of Dirac structures will find their places in the final expressions for the form factors. Cutting rules were also adopted in \cite{MarquesdeCarvalho:1999bqs}, but only the coefficients of the Dirac structures with the highest dimension were used to extract the form factors.

For the processes of $\Lambda_{b}\to\Lambda_{c}$ and $\Xi_{b}\to\Xi_{c}$, the leading order
contributions from dimension-3 and dimension-5 operators are respectively proportional to the mass of the light quark and the mass of the strange quark. For the former, these contributions can be neglected. Therefore, we will set the process $\Xi_{b}\to\Xi_{c}$ as default in the following analysis. The corresponding results of $\Lambda_{b}\to\Lambda_{c}$ will also be shown when appropriate.
 When performing the numerical analysis, the Wilson coefficients are calculated with perturbative QCD, thus we will employ the $\overline{{\rm MS}}$
scheme for the quark masses.  If we take the heavy quark limit the HQET sum rules results can be reproduced, as can be seen
in \cite{Shuryak:1981fza,Zhao:2020wbw,MarquesdeCarvalho:1999bqs}.

The rest of this paper is arranged as follows. In Sec. II,
we will discuss the two-point correlation functions to evaluate the
pole residues of $\Xi_{Q}$ and $\Lambda_{Q}$ for completeness. In Sec. III, we will
investigate the three-point correlation functions to arrive at the
analytical results of the form factors. Numerical results for the form factors and their phenomenological applications
will be shown in Sec. IV. In this section, we will also compare our results with other theoretical predictions
 and the experimental data to test the validity
of our  calculation. We conclude this paper in the last section.

\section{The two-point correlation functions and pole residues}
\label{sec:2p}

Pole residues of heavy baryons have also been investigated in the literature \cite{Wang:2010fq,Wang:2020mxk}. For completeness, we still briefly describe the calculation of the two-point correlation functions in this section.

To construct the correlation function, one should choose the appropriate  interpolating currents for $\Lambda_{Q}$ and $\Xi_{Q}$. As the isospin of the diquark $[ud]$ in the baryon $\Lambda_{(c,b)}$ is 0, we adopt the following interpolating currents in our calculation:
\begin{align}
J_{\Lambda_{Q}} & =\epsilon_{abc}(u_{a}^{T}C\gamma_{5}d_{b})Q_{c},\nonumber \\
J_{\Xi_{Q}} & =\epsilon_{abc}(q_{a}^{T}C\gamma_{5}s_{b})Q_{c},\label{eq:interpolating_current}
\end{align}
where $Q=b$ or $c$, $q=u$ or $d$, the color indices are denoted
by $a,b,c$ and $C$ is the charge conjugate matrix. The two-point
correlation function is defined by
\begin{equation}
\Pi(p)=i\int d^{4}x\ e^{ip\cdot x}\langle0|T\{J(x)\bar{J}(0)\}|0\rangle.
\label{eq:2p_correlator}
\end{equation}
 On the hadronic side, one can insert the complete set of hadronic
states to write the above correlation function as
\begin{equation}
\Pi^{{\rm had}}(p)=\lambda_{+}^{2}\frac{\slashed p+M_{+}}{M_{+}^{2}-p^{2}}+\lambda_{-}^{2}\frac{\slashed p-M_{-}}{M_{-}^{2}-p^{2}}+\cdots,\label{eq:correlator_had}
\end{equation}
where we have also considered the contribution from the negative parity
baryon, $M_{\pm}$ ($\lambda_{\pm}$) stand for the masses (the pole
residues) of the positive and negative parity baryons.

\begin{figure}[!]
\includegraphics[width=0.8\columnwidth]{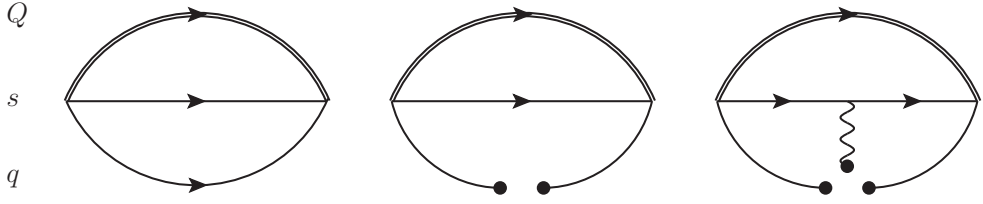}
\caption{Only 3 nonzero diagrams survive for the two-point correlation function of $\Xi_{Q}$, if we consider the contributions up to dim-5 and neglect those from dim-4. The double lines denote the heavy quarks and the dots stand for the condensates.}
\label{fig:diagrams_2p}
\end{figure}

On the QCD side, we evaluate the  correlation function in Eq. (\ref{eq:2p_correlator})  following
the OPE technique. Since the contributions from
gluon condensate are small \cite{Zhao:2020wbw}, one can only consider
the contributions from dim-0,3,5 operators and the corresponding
nonzero diagrams can be found in Fig. \ref{fig:diagrams_2p}. The result can be formally written as
\begin{equation}
\Pi^{{\rm QCD}}(p)=A(p^{2})\slashed p+B(p^{2}),
\label{eq:2p_correlator_QCD}
\end{equation}
where the coefficients $A$ and $B$ can be written in terms of the dispersion integrals
\begin{equation}
A(p^{2})=\int ds\ \frac{\rho^{A}(s)}{s-p^{2}},\quad B(p^{2})=\int ds\ \frac{\rho^{B}(s)}{s-p^{2}}.\label{eq:rho_A_rho_B}
\end{equation}

Taking advantage of the quark-hadron duality assumption and employing the Borel transform, the QCD sum rule for the pole residue of $1/2^{+}$ baryon is given by
\begin{align}
(M_{+}+M_{-})\lambda_{+}^{2}\exp(-M_{+}^{2}/T_{+}^{2}) & =\int_{m_{Q}^{2}}^{s_{+}}ds\ (M_{-}\rho^{A}+\rho^{B})\exp(-s/T_{+}^{2}),\label{eq:2p_sum_rule}
\end{align}
where $T_{+}^{2}$ is the Borel parameters and $s_{+}$ is the
continuum threshold parameters. Differentiating the equation in Eq. (\ref{eq:2p_sum_rule}) with respect to $-1/T_{+}^{2}$, one can arrive at the sum rule for the mass of $1/2^{+}$ baryon
\begin{equation}
M_{+}^{2}=\frac{\int_{m_{Q}^{2}}^{s_{+}}ds\ (M_{-}\rho^{A}+\rho^{B})\ s\ \exp(-s/T_{+}^{2})}{\int_{m_{Q}^{2}}^{s_{+}}ds\ (M_{-}\rho^{A}+\rho^{B})\ \exp(-s/T_{+}^{2})}.
\label{eq:mass_sum_rule}
\end{equation}
In practice, Eq. (\ref{eq:mass_sum_rule}) is used to test the sum rule in Eq. (\ref{eq:2p_sum_rule}).

\section{Three-point correlation functions and form factors}
\label{sec:3p}

For the  $\Lambda_{b}\to\Lambda_{c}$ and $\Xi_{b}\to\Xi_{c}$ transition form factors, we take advantage of  the following simpler parametrization in this section:
\begin{align}
&\langle{\cal B}_{2}(p_{2},s_{2})|\bar{c}\gamma_{\mu}(1-\gamma_{5})b|{\cal B}_{1}(p_{1},s_{1})\rangle \nonumber \\ =&\bar{u}(p_{2},s_{2})\left[\frac{p_{1\mu}}{M_{1}}F_{1}(q^{2})
+\frac{p_{2\mu}}{M_{2}}F_{2}(q^{2})+\gamma_{\mu}F_{3}(q^{2})\right]u(p_{1},s_{1})\nonumber \\
 -&\bar{u}(p_{2},s_{2})\left[\frac{p_{1\mu}}{M_{1}}G_{1}(q^{2})
 +\frac{p_{2\mu}}{M_{2}}G_{2}(q^{2})+\gamma_{\mu}G_{3}(q^{2})\right]\gamma_{5}u(p_{1},s_{1}),\label{eq:parametrization_simple}
\end{align}
where ${\cal B}_{1,2}$ denote $(\Lambda,\Xi)_{b,c}$. The form factors  $F_{i},G_{i}$ are related to $f_{i},g_{i}$ defined in (\ref{eq:parametrization_standard}) through
\begin{align}
 & F_{1}=f_{2}+f_{3},\quad F_{2}=\frac{M_{2}}{M_{1}}(f_{2}-f_{3}),\quad F_{3}=f_{1}-\frac{M_{1}+M_{2}}{M_{1}}f_{2};\nonumber \\
 & G_{1}=g_{2}+g_{3},\quad G_{2}=\frac{M_{2}}{M_{1}}(g_{2}-g_{3}),\quad G_{3}=g_{1}+\frac{M_{1}-M_{2}}{M_{1}}g_{2}.
\end{align}
Thus at the leading power of $1/m_{Q}$ in HQET, the form factors defined in Eq. (\ref{eq:parametrization_simple})
satisfies \cite{Manohar:2000dt}
\begin{equation}
F_{1}=F_{2}=G_{1}=G_{2}=0,\qquad F_{3}(\omega)=G_{3}(\omega)=\zeta(\omega),
\end{equation}
with
\begin{equation}
\zeta(\omega=1)=1,
\end{equation}
where $\omega\equiv v_{1}\cdot v_{2}=(p_{1}\cdot p_{2})/(M_{1}M_{2})$.

As mentioned before, we will take $\Xi_{b}\to\Xi_{c}$ transition as the default process to illustrate our method. The correlation function for $\Xi_{b}\to\Xi_{c}$ transition is defined
as
\begin{equation}
\Pi_{\mu}^{V,A}(p_{1}^{2},p_{2}^{2},q^{2})=i^{2}\int d^{4}xd^{4}y\ e^{-ip_{1}\cdot x+ip_{2}\cdot y}\langle0|T\{J_{\Xi_{c}}(y)(V_{\mu},A_{\mu})(0)\bar{J}_{\Xi_{b}}(x)\}|0\rangle,
\end{equation}
where $V_{\mu}(A_{\mu})=\bar{c}\gamma_{\mu}(\gamma_{\mu}\gamma_{5})b$
is the vector (axial-vector) current for $b\to c$ weak decay. The
interpolating currents for initial and final states can be found in
Eqs. (\ref{eq:interpolating_current}).

Following the stardard steps of QCD sum rules, the correlation function
will be calculated at hadronic level and QCD level. At the hadronic
level, after inserting the complete set of initial and final states,
the vector current correlation function can be written as
\begin{equation}
\Pi_{\mu}^{V,{\rm had}}(p_{1}^{2},p_{2}^{2},q^{2})=\lambda_{f}\lambda_{i}\frac{(\slashed p_{2}+M_{2})(\frac{p_{1\mu}}{M_{1}}F_{1}+\frac{p_{2\mu}}{M_{2}}F_{2}+\gamma_{\mu}F_{3})(\slashed p_{1}+M_{1})}{(p_{2}^{2}-M_{2}^{2})(p_{1}^{2}-M_{1}^{2})}+\cdots,\label{eq:correlator_had_old}
\end{equation}
where $\lambda_{i(f)}=\lambda_{\Xi_{b(c)}}$, $F_{i}$ are form factors defined in Eq. (\ref{eq:parametrization_simple}),
$M_{1,2}$ are the masses of initial and final states and the ellipsis
stands for the contribution from higher resonances and continuum spectra.
It is clear that there are 12 Dirac structures, but only 3 form
factors to be determined in Eq. (\ref{eq:correlator_had_old}). For each form factor, there are 4 Dirac
structures available. Furthermore, it is very likely that these  different
Dirac structures give rise to very different results since only the LO results are considered. To eliminate these ambiguities, we consider again the contributions from the negative parity baryons, which have been swept into the ellipsis in Eq. (\ref{eq:correlator_had_old}). After that, the vector current
correlation function can be rewritten as
\begin{eqnarray}
\Pi_{\mu}^{V,{\rm had}}(p_{1}^{2},p_{2}^{2},q^{2}) & = & \lambda_{f}^{+}\lambda_{i}^{+}\frac{(\slashed p_{2}+M_{2}^{+})(\frac{p_{1\mu}}{M_{1}^{+}}F_{1}^{++}+\frac{p_{2\mu}}{M_{2}^{+}}F_{2}^{++}+\gamma_{\mu}F_{3}^{++})(\slashed p_{1}+M_{1}^{+})}{(p_{2}^{2}-M_{2}^{+2})(p_{1}^{2}-M_{1}^{+2})}\nonumber \\
 & + & \lambda_{f}^{+}\lambda_{i}^{-}\frac{(\slashed p_{2}+M_{2}^{+})(\frac{p_{1\mu}}{M_{1}^{-}}F_{1}^{+-}+\frac{p_{2\mu}}{M_{2}^{+}}F_{2}^{+-}+\gamma_{\mu}F_{3}^{+-})(\slashed p_{1}-M_{1}^{-})}{(p_{2}^{2}-M_{2}^{+2})(p_{1}^{2}-M_{1}^{-2})}\nonumber \\
 & + & \lambda_{f}^{-}\lambda_{i}^{+}\frac{(\slashed p_{2}-M_{2}^{-})(\frac{p_{1\mu}}{M_{1}^{+}}F_{1}^{-+}+\frac{p_{2\mu}}{M_{2}^{-}}F_{2}^{-+}+\gamma_{\mu}F_{3}^{-+})(\slashed p_{1}+M_{1}^{+})}{(p_{2}^{2}-M_{2}^{-2})(p_{1}^{2}-M_{1}^{+2})}\nonumber \\
 & + & \lambda_{f}^{-}\lambda_{i}^{-}\frac{(\slashed p_{2}-M_{2}^{-})(\frac{p_{1\mu}}{M_{1}^{-}}F_{1}^{--}+\frac{p_{2\mu}}{M_{2}^{-}}F_{2}^{--}+\gamma_{\mu}F_{3}^{--})(\slashed p_{1}-M_{1}^{-})}{(p_{2}^{2}-M_{2}^{-2})(p_{1}^{2}-M_{1}^{-2})}\nonumber \\
 & + & \cdots.\label{eq:correlator_had_new}
\end{eqnarray}
In Eq. (\ref{eq:correlator_had_new}), $M_{1(2)}^{+(-)}$ denotes
the masses of initial (final) positive (negative) parity baryons,
and $F_{1}^{-+}$ is the form factor $F_{1}$
with the negative-parity final state and the positive-parity initial
state, and so forth. To arrive at Eq. (\ref{eq:correlator_had_new})
, we have also adopted the definitions of pole residues for positive
and negative parity baryons
\begin{align}
\langle0|J_{+}|{\cal B}_{+}(p,s)\rangle & =\lambda_{+}u(p,s),\nonumber \\
\langle0|J_{+}|{\cal B}_{-}(p,s)\rangle & =(i\gamma_{5})\lambda_{-}u(p,s),\label{eq:pole_residue}
\end{align}
and the following conventions for the form factors $F_{i}^{\pm\pm}$:
\begin{align}
\langle{\cal B}_{f}^{+}(p_{2},s_{2})|V_{\mu}|{\cal B}_{i}^{+}(p_{1},s_{1})\rangle & =\bar{u}(p_{2},s_{2})[\frac{p_{1\mu}}{M_{1}^{+}}F_{1}^{++}+\frac{p_{2\mu}}{M_{2}^{+}}F_{2}^{++}+\gamma_{\mu}F_{3}^{++}]u(p_{1},s_{1}),\nonumber \\
\langle{\cal B}_{f}^{+}(p_{2},s_{2})|V_{\mu}|{\cal B}_{i}^{-}(p_{1},s_{1})\rangle & =\bar{u}(p_{2},s_{2})[\frac{p_{1\mu}}{M_{1}^{-}}F_{1}^{+-}+\frac{p_{2\mu}}{M_{2}^{+}}F_{2}^{+-}+\gamma_{\mu}F_{3}^{+-}](i\gamma_{5})u(p_{1},s_{1}),\nonumber \\
\langle{\cal B}_{f}^{-}(p_{2},s_{2})|V_{\mu}|{\cal B}_{i}^{+}(p_{1},s_{1})\rangle & =\bar{u}(p_{2},s_{2})(i\gamma_{5})[\frac{p_{1\mu}}{M_{1}^{+}}F_{1}^{-+}+\frac{p_{2\mu}}{M_{2}^{-}}F_{2}^{-+}+\gamma_{\mu}F_{3}^{-+}]u(p_{1},s_{1}),\nonumber \\
\langle{\cal B}_{f}^{-}(p_{2},s_{2})|V_{\mu}|{\cal B}_{i}^{-}(p_{1},s_{1})\rangle & =\bar{u}(p_{2},s_{2})(i\gamma_{5})[\frac{p_{1\mu}}{M_{1}^{-}}F_{1}^{--}+\frac{p_{2\mu}}{M_{2}^{-}}F_{2}^{--}+\gamma_{\mu}F_{3}^{--}](i\gamma_{5})u(p_{1},s_{1}).
\end{align}
In Eq. (\ref{eq:pole_residue}), $J_{+}$ can be found in Eqs. (\ref{eq:interpolating_current}),
and $\lambda_{+(-)}$ is the pole residue for the positive (negative)
parity baryon.

\begin{figure}[!]
\includegraphics[width=0.8\columnwidth]{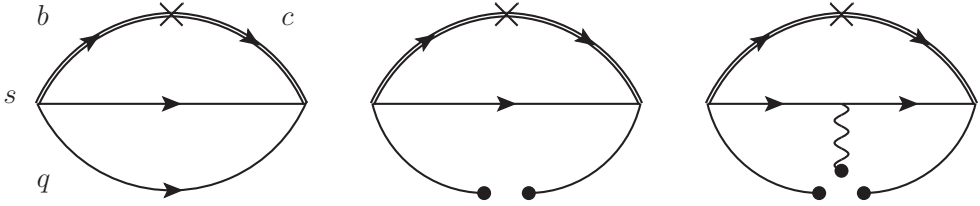}
\caption{Only 3 nonzero diagrams survive for the three-point correlation function of $\Xi_{b}$ decaying into $\Xi_{c}$, if we consider the contributions up to dim-5 and neglect those from dim-4. The double lines denote the heavy quarks, the dots stand for the condensates, and the cross marks are vertices of weak interaction.}
\label{fig:diagrams_3p}
\end{figure}

At the QCD level, there are three diagrams to be considered up to dim-5, as can be seen in Fig. \ref{fig:diagrams_3p}. \footnote{As can be seen in \cite{Shi:2019hbf}, the contributions from gluon condensate are  small, thereby we do not consider them in this work.} For practical purpose, the correlation function is expressed as a double
dispersion relation
\begin{equation}
\Pi_{\mu}^{V,{\rm QCD}}(p_{1}^{2},p_{2}^{2},q^{2})=\int^{\infty}ds_{1}\int^{\infty}ds_{2}\frac{\rho_{\mu}^{V,{\rm QCD}}(s_{1},s_{2},q^{2})}{(s_{1}-p_{1}^{2})(s_{2}-p_{2}^{2})},\label{eq:3p_correlator_QCD}
\end{equation}
with $\rho_{\mu}^{V,{\rm QCD}}(s_{1},s_{2},q^{2})$ being the spectral
function, which can be obtained by applying Cutkosky cutting rules. Based on the assumption of quark-hadron duality,
the sum of the four pole terms in Eq. (\ref{eq:correlator_had_new})
should be equal to
\begin{equation}
\int^{s_{1}^{0}}ds_{1}\int^{s_{2}^{0}}ds_{2}\frac{\rho_{\mu}^{V,{\rm QCD}}(s_{1},s_{2},q^{2})}{(s_{1}-p_{1}^{2})(s_{2}-p_{2}^{2})}\equiv\Pi_{\mu}^{V,{\rm pole}},
\end{equation}
where $s_{1(2)}^{0}$ is the continuum threshold parameter for the
initial (final) baryon. $\Pi_{\mu}^{V,{\rm pole}}$ can be formally
written as
\begin{equation}
\Pi_{\mu}^{V,{\rm pole}}=\sum_{i=1}^{12}A_{i}e_{i\mu},
\label{eq:correlator_pole_formal}
\end{equation}
where, we have defined
\begin{eqnarray}
(e_{1,2,3,4})_{\mu} & = & \{\slashed p_{2},1\}\times\{p_{1\mu}\}\times\{\slashed p_{1},1\},\nonumber \\
(e_{5,6,7,8})_{\mu} & = & \{\slashed p_{2},1\}\times\{p_{2\mu}\}\times\{\slashed p_{1},1\},\nonumber \\
(e_{9,10,11,12})_{\mu} & = & \{\slashed p_{2},1\}\times\{\gamma_{\mu}\}\times\{\slashed p_{1},1\}.\label{eq:e_i_mu}
\end{eqnarray}
By equating Eq. (\ref{eq:correlator_had_new}) with Eq. (\ref{eq:correlator_pole_formal}), one can obtain 12 equations. Solving these equations, one can obtain these 12 form factors $F_{i}^{\pm,\pm}$, including the following 3 expressions for $F_{i}^{++}$:
\begin{eqnarray}
\frac{\lambda_{i}^{+}\lambda_{f}^{+}(F_{1}^{++}/M_{1}^{+})}{(p_{1}^{2}-M_{1}^{+2})(p_{2}^{2}-M_{2}^{+2})} & = & \frac{\{M_{1}^{-}M_{2}^{-},M_{2}^{-},M_{1}^{-},1\}.\{A_{1},A_{2},A_{3},A_{4}\}}{(M_{1}^{+}+M_{1}^{-})(M_{2}^{+}+M_{2}^{-})},\nonumber \\
\frac{\lambda_{i}^{+}\lambda_{f}^{+}(F_{2}^{++}/M_{2}^{+})}{(p_{1}^{2}-M_{1}^{+2})(p_{2}^{2}-M_{2}^{+2})} & = & \frac{\{M_{1}^{-}M_{2}^{-},M_{2}^{-},M_{1}^{-},1\}.\{A_{5},A_{6},A_{7},A_{8}\}}{(M_{1}^{+}+M_{1}^{-})(M_{2}^{+}+M_{2}^{-})},\nonumber \\
\frac{\lambda_{i}^{+}\lambda_{f}^{+}F_{3}^{++}}{(p_{1}^{2}-M_{1}^{+2})(p_{2}^{2}-M_{2}^{+2})} & = & \frac{\{M_{1}^{-}M_{2}^{-},M_{2}^{-},M_{1}^{-},1\}.\{A_{9},A_{10},A_{11},A_{12}\}}{(M_{1}^{+}+M_{1}^{-})(M_{2}^{+}+M_{2}^{-})}.
\end{eqnarray}
Borel transform the above equations to suppress the contributions from higher resonances and continuum spectra to arrive at:
\begin{align}
\lambda_{i}^{+}\lambda_{f}^{+}(F_{1}^{++}/M_{1}^{+})\exp\left(-\frac{M_{1}^{+2}}{T_{1}^{2}}-\frac{M_{2}^{+2}}{T_{2}^{2}}\right) & =\frac{\{M_{1}^{-}M_{2}^{-},M_{2}^{-},M_{1}^{-},1\}.\{{\cal B}A_{1},{\cal B}A_{2},{\cal B}A_{3},{\cal B}A_{4}\}}{(M_{1}^{+}+M_{1}^{-})(M_{2}^{+}+M_{2}^{-})},\nonumber \\
\lambda_{i}^{+}\lambda_{f}^{+}(F_{2}^{++}/M_{2}^{+})\exp\left(-\frac{M_{1}^{+2}}{T_{1}^{2}}-\frac{M_{2}^{+2}}{T_{2}^{2}}\right) & =\frac{\{M_{1}^{-}M_{2}^{-},M_{2}^{-},M_{1}^{-},1\}.\{{\cal B}A_{5},{\cal B}A_{6},{\cal B}A_{7},{\cal B}A_{8}\}}{(M_{1}^{+}+M_{1}^{-})(M_{2}^{+}+M_{2}^{-})},\nonumber \\
\lambda_{i}^{+}\lambda_{f}^{+}F_{3}^{++}\exp\left(-\frac{M_{1}^{+2}}{T_{1}^{2}}-\frac{M_{2}^{+2}}{T_{2}^{2}}\right) & =\frac{\{M_{1}^{-}M_{2}^{-},M_{2}^{-},M_{1}^{-},1\}.\{{\cal B}A_{9},{\cal B}A_{10},{\cal B}A_{11},{\cal B}A_{12}\}}{(M_{1}^{+}+M_{1}^{-})(M_{2}^{+}+M_{2}^{-})},\label{eq:Fi_plus_plus}
\end{align}
where ${\cal B}A_{i}\equiv{\cal B}_{T_{1}^{2},T_{2}^{2}}A_{i}$ are
doubly Borel transformed coefficients, and $T_{1,2}^{2}$ are the Borel mass parameters.

To obtain the coefficients $A_{i}$ in Eq. (\ref{eq:correlator_pole_formal}), one can project Eq. (\ref{eq:correlator_pole_formal}) onto 12 Dirac structures. Specifically, multiplying by $e_{j}^{\mu}$
and then taking traces on both sides of Eq. (\ref{eq:correlator_pole_formal}),
one can arrive at the following 12 linear equations:
\begin{equation}
B_{j}\equiv{\rm Tr}[\Pi_{\mu}^{V,{\rm pole}}e_{j}^{\mu}]={\rm Tr}\left[\left(\sum_{i=1}^{12}A_{i}e_{i\mu}\right)e_{j}^{\mu}\right],\quad j=1,...,12.\label{eq:projection}
\end{equation}
Solving these equations, one can obtain the expressions of $A_{i}$ given that it is easy to write down $\Pi_{\mu}^{V,{\rm pole}}$.

\section{Numerical results and phenomenological applications}

In our numerical calculations, the condensate parameters are taken
as \cite{Colangelo:2000dp}: $\langle\bar{q}q\rangle=-(0.24\pm0.01\ {\rm GeV})^{3}$,
$\langle\bar{q}g_{s}\sigma Gq\rangle=m_{0}^{2}\langle\bar{q}q\rangle$,
$m_{0}^{2}=(0.8\pm0.2)\ {\rm GeV}^{2}$, where the renormalization
scale is taken at $\mu=1$ GeV. 
 In this work, we will use $\overline{{\rm MS}}$ masses for quarks
unless otherwise stated. When dealing with the two-point
correction functions for bottom baryons, we take the renormalization
scale at $\mu=m_{b}$, while for charmed baryons, $\mu=m_{c}$. For the three-point
correction functions of a bottom baryon decaying into a charmed baryon, we take $\mu=m_{b}$.
The following quark masses are used \cite{Tanabashi:2018oca}:
\begin{align}
 & m_{b}(m_{b})=4.18\pm0.03\ {\rm GeV},\quad m_{c}(m_{b})=1.02\pm0.02\ {\rm GeV},\quad m_{s}(m_{b})=(0.082\pm0.010)\ {\rm GeV},\nonumber \\
 & m_{c}(m_{c})=1.27\pm0.02\ {\rm GeV},\quad m_{s}(m_{c})=0.103\pm0.012\ {\rm GeV}.\label{eq:quark_mass}
\end{align}
Since we are considering the LO calculation of QCDSR, when arriving
at the above masses, it would be enough to adopt the following one-loop
evolution equation
\begin{equation}
m_{q}(\mu_{0})\left(\frac{\log(\mu_{0}/\Lambda_{{\rm QCD}}^{(n_{f})})}{\log(\mu/\Lambda_{{\rm QCD}}^{(n_{f})})}\right)^{4/\beta_{0}}
\end{equation}
for $m_{c}(m_{b})$, $m_{s}(m_{b})$ and $m_{s}(m_{c})$. In the above
equation, $\beta_{0}=11-(2/3)n_{f}$ with $n_{f}$ the number of active
flavors, and $\Lambda_{{\rm QCD}}^{(4)}=170\ {\rm MeV}$ has been
used in Eqs. (\ref{eq:quark_mass}). In the following, $\Lambda_{{\rm QCD}}^{(3)}=223\ {\rm MeV}$
will also be used. These two values for $\Lambda_{{\rm QCD}}$ are
obtained by demanding the results for $\alpha_{s}$ at the LO can
reproduce the corresponding results at the NLO \cite{Buras:1998raa} \footnote{In fact, $\Lambda_{{\rm QCD}}^{(4)}=147\ {\rm MeV}$
and $\Lambda_{{\rm QCD}}^{(4)}=194\ {\rm MeV}$ are respectively obtained
at $\mu=m_{b}$ and $\mu=m_{c}$, while $\Lambda_{{\rm QCD}}^{(3)}=223\ {\rm MeV}$
is obtained at $\mu=m_{c}$. Mean value
is adopted for $\Lambda_{{\rm QCD}}^{(4)}$ in our QCDSR calculation.}.

\subsection{The two-point correlation function}

In this work, we will also consider the leading logarithm (LL) approximation
for the pole residues and masses of baryons. According to \cite{Ioffe:1981kw},
the Wilson coefficients of the local operators that we derived in Sec. \ref{sec:2p} should
be multiplied by an evolution factor
\begin{equation}
\left(\frac{\log(\mu_{0}/\Lambda_{{\rm QCD}}^{(n_{f})})}{\log(\mu/\Lambda_{{\rm QCD}}^{(n_{f})})}\right)^{2\gamma_{J}-\gamma_{O}},\label{eq:LL_correction}
\end{equation}
where $\gamma_{J}$ is the anomalous dimension of the current $J$
in Eq. (\ref{eq:interpolating_current}), and $\gamma_{O}$ is that
of the local operator in the OPE. Following \cite{Ioffe:1981kw}, we will also only consider the LL corrections for
the perturbative and quark condensate contributions. $\mu_{0}$ is
the renormalization scale of the low-energy limit \cite{Ioffe:1981kw}, which
is roughly at $1\ {\rm GeV}$, and $\mu\sim m_{Q}$ is the renormalization
scale that we choose for the physical quantities of interest. For the interpolating current given in Eq. (\ref{eq:interpolating_current}),
the corresponding anomalous dimension is $\gamma_{J}=-1/\beta_{0}$ \cite{Ovchinnikov:1991mu}. The anomalous dimension
for $\bar{\psi}\psi$ is given by $\gamma_{\bar{\psi}\psi}=4/\beta_{0}$.

\begin{figure}[!]
\includegraphics[width=\columnwidth]{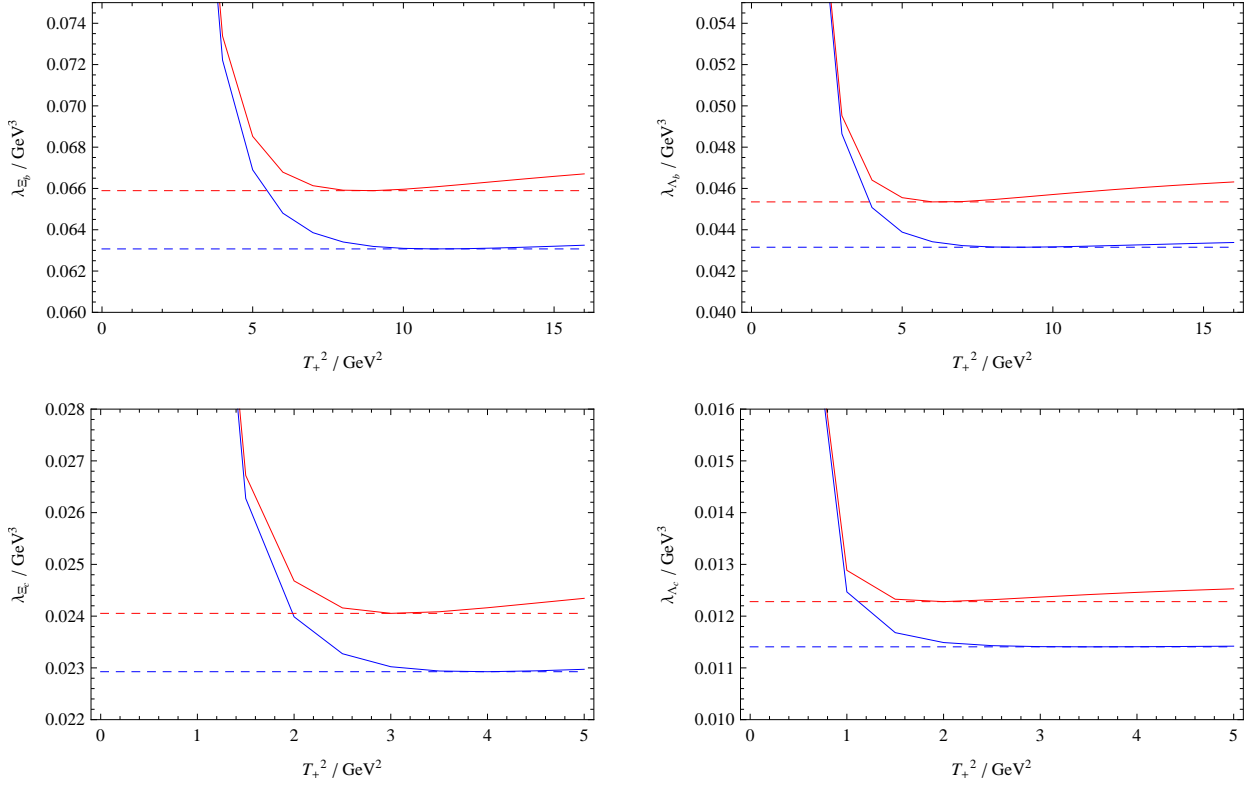}
\caption{
The pole residues of $\Xi_{b}$ (top left), $\Lambda_{b}$ (top right), $\Xi_{c}$ (bottom left) and $\Lambda_{c}$ (bottom right) as functions of the Borel parameters $T_{+}^{2}$. The blue and red curves correspond to optimal and suboptimal choices for $s_{+}$. The extreme points of these curves correspond to the optimal choices of $T_{+}^{2}$ on these curves. The explicit values for these $s_{+}$ and $T_{+}^{2}$ can be found in Table \ref{Tab:pole_residues}.
}
\label{fig:pole_residues}
\end{figure}

\begin{table}
\caption{The predictions of the pole residues and masses. For comparison, the
experimental values of heavy baryons are also shown. Optimal and suboptimal
$s_{+}$ and $T_{+}^{2}$ are given simultaneously. The central values
for the pole residues and masses are taken at optimal $s_{+}$
and $T_{+}^{2}$, and the values at suboptimal $s_{+}$ and $T_{+}^{2}$
provide the error estimates. The results of bottom and charmed baryons are respectively obtained at the renormalization scale $\mu=m_{b}$ and  $\mu=m_{c}$, respectively.}
\label{Tab:pole_residues} %
\begin{tabular}{c|c|c|c|c}
\hline
 & $(s_{+}/{\rm GeV}^{2},\ T_{+}^{2}/{\rm GeV}^{2})$  & $\lambda_{+}/{\rm GeV}^{3}$  & $M_{+}/{\rm GeV}$  & $M_{+}^{{\rm exp}}/{\rm GeV}$\tabularnewline
\hline
$\Xi_{b}$  & $(6.20^{2},11),\ (6.25^{2},9)$  & $0.0631\pm0.0028$  & $5.791\pm0.009$  & $5.793$\tabularnewline
$\Xi_{c}$  & $(2.85^{2},4.0),\ (2.90^{2},3.0)$  & $0.0229\pm0.0011$  & $2.471\pm0.007$  & $2.468$\tabularnewline
\hline
$\Lambda_{b}$  & $(5.95^{2},9)\ (6.00^{2},6)$  & $0.0432\pm0.0022$  & $5.622\pm0.010$  & $5.620$\tabularnewline
$\Lambda_{c}$  & $(2.50^{2},3.5),\ (2.55^{2},2.0)$  & $0.0114\pm0.0009$  & $2.286\pm0.005$  & $2.286$\tabularnewline
\hline
\end{tabular}
\end{table}

Using the sum rule in Eq. (\ref{eq:2p_sum_rule}), we can determine
the pole residues for $(\Lambda,\Xi)_{b,c}$. The
pole residues as functions of the Borel parameter $T_{+}^{2}$ are
given in Fig. \ref{fig:pole_residues}, from which, we arrive at our
predictions of the pole residues and masses for $(\Lambda,\Xi)_{b,c}$ in Table
\ref{Tab:pole_residues}. Some comments are in order.

\begin{itemize}
\item It can be seen that our predictions for the masses of $(\Lambda,\Xi)_{b,c}$
are in very good agreement with the experimental results. Presumably
it is due to the overwhelming contribution from perturbative diagram,
since the second and third diagrams in Fig. \ref{fig:diagrams_2p}
are proportional to the mass of light quark.
\item It turns out that the LL corrections for dim-0,3 are respectively $16\%$, $54\%$ for the bottom baryons, and $3\%$,
$10\%$ for the charmed baryons. Although the corrections for dim-3
is large, it does not play an important role because of the fact stated in the last item.
\end{itemize}

\subsection{The three-point correlation function}

To access the numerical results for the form factors of $\Xi_{b}\to\Xi_{c}$,
firstly we need to find out the optimal choices for the threshold
parameters $s_{1,2}^{0}$ and the Borel masses $T_{1,2}^{2}$. For
the former, we just borrow them from the corresponding two-point
correlation functions. The optimal values for $s_{1}^{0}$ and $s_{2}^{0}$
are respectively $(6.20\ {\rm GeV})^{2}$ and $(2.85\ {\rm GeV})^{2}$,
as can be seen from Table \ref{Tab:pole_residues}. Then we scan the
$T_{1}^{2}-T_{2}^{2}$ plane to determine the optimal Borel region.

Note that we also consider the LL resummation for the form factors. Since
the anomalous dimensions for the vector current and axial-vector current
 vanish, the Wilson coefficients of the local operators that we derived
in Sec. \ref{sec:3p} should be multiplied by the same evolution factor as in
Eq. (\ref{eq:LL_correction}). One more thing should be addressed: the pole residue for $\Xi_{c}$ ($\Lambda_{c}$) in Table \ref{Tab:pole_residues}, which is evaluated at $\mu=m_{c}$, should be evolved to the scale of $\mu=m_{b}$.

However, we fail to find stable regions like the cases of the two-point
correlation functions. To find relatively optimal regions on $T_{1}^{2}-T_{2}^{2}$
plane, the following criteria is to be employed:
\begin{itemize}

\item Pole dominance. We demand
\begin{align}
r_{1} & \equiv\frac{\int^{s_{1}^{0}}ds_{1}\int^{s_{2}^{0}}ds_{2}}{\int^{\infty}ds_{1}\int^{s_{2}^{0}}ds_{2}}\gtrsim0.5,\nonumber \\
r_{2} & \equiv\frac{\int^{s_{1}^{0}}ds_{1}\int^{s_{2}^{0}}ds_{2}}{\int^{s_{1}^{0}}ds_{1}\int^{\infty}ds_{2}}\gtrsim0.5,\label{eq:pole_dominance}
\end{align}
which can be viewed as the pole dominance criteria for $\Xi_{b}$ ($\Lambda_{b}$)
channel and $\Xi_{c}$ ($\Lambda_{c}$) channel, respectively.

\item OPE convergence. It can be achieved by demanding that dim-5/Total
should be small enough.

\end{itemize}

\begin{figure}[!]
\includegraphics[width=0.9\columnwidth]{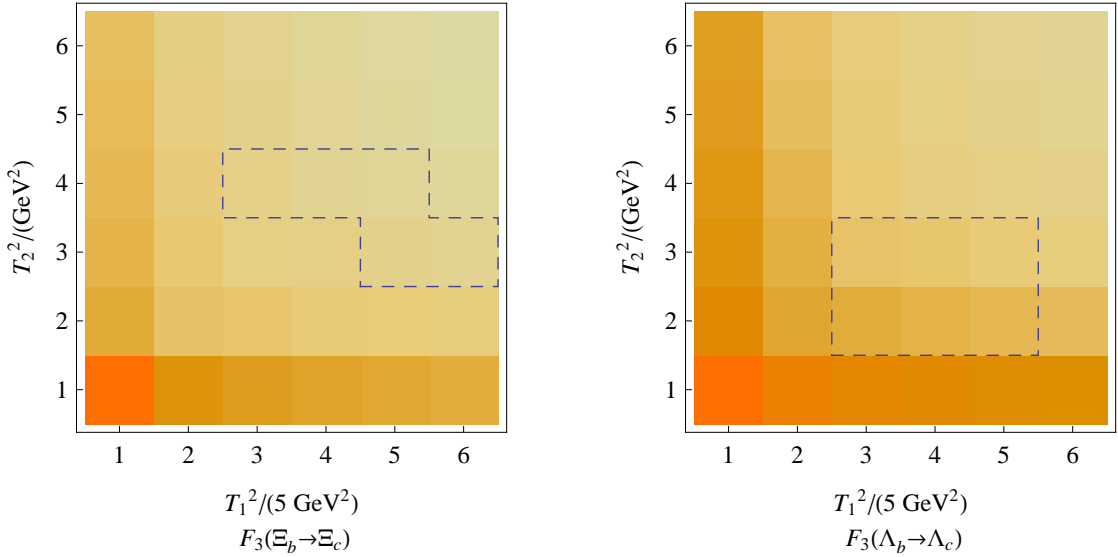}
\caption{$F_{3}^{\Xi_{b}\to\Xi_{c}}$ and $F_{3}^{\Lambda_{b}\to\Lambda_{c}}$
at $q^{2}=0$ as functions of the Borel parameters $T_{1}^{2}$ and
$T_{2}^{2}$, where $T_{1}^{2}$ and $T_{2}^{2}$ are taken as free
parameters. The larger the form factors, the
darker the color. The preferred Borel regions are enclosed by the
dashed contours.}
\label{fig:Borel}
\end{figure}

Complying with the above criteria, we arrive at the relatively optimal
regions on the $T_{1}^{2}-T_{2}^{2}$ plane for $\Xi_{b}\to\Xi_{c}$,
which are enclosed by the dashed contours, as plotted in Fig. \ref{fig:Borel}.
The specific values of $r_{1,2}$ and dim-5/Total in these
regions can be found in Table \ref{Tab:criteria}. Since $F_{1,2}$
is small, the definitions in Eqs. (\ref{eq:pole_dominance}) may
be ill-defined, so we only give the selected regions for $F_{3}$,
and just assume that the same regions are also applied to $F_{1,2}$.
Similar regions can be obtained for $G_{3}$ and same assumption is
applied to $G_{1,2}$.

For $\Lambda_{b}\to\Lambda_{c}$, the contributions from dim-3,5 are
neglected because they are proportional to the mass of the light quark, thereby the second criterion does not work. However, one can see that in the
selected regions for $\Xi_{b}\to\Xi_{c}$, $T_{1}^{2}\sim{\cal O}(m_{b}^{2})$
and $T_{2}^{2}\sim{\cal O}(m_{c}^{2})$, so it is plausible that similar pattern should also hold for $\Lambda_{b}\to\Lambda_{c}$. By constraining
$T_{1}^{2}\in[15,25]\ {\rm GeV}^{2}$, $T_{2}^{2}\in[2,4]\ {\rm GeV}^{2}$
and also considering the first criterion above, the Borel region for $\Lambda_{b}\to\Lambda_{c}$
can also be determined, as can be seen in Fig. \ref{fig:Borel}.

\begin{table}
\caption{ The quantitative criteria of the pole dominance and OPE convergence.}
\label{Tab:criteria} %
\begin{tabular}{c|c||c|c}
\hline
$\Xi_{b}\to\Xi_{c}$ & $F_{3}(0)$  & $\Lambda_{b}\to\Lambda_{c}$ & $F_{3}(0)$ \tabularnewline
\hline
$r_{1}$  & $>59\%$  & $r_{1}$  & $>56\%$ \tabularnewline
$r_{2}$  & $>50\%$  & $r_{2}$  & $>39\%$ \tabularnewline
dim-5/total  & $<10\%$  & dim-5/total  & - -\tabularnewline
\hline
\end{tabular}
\end{table}

After all the parameters are fixed, central values and uncertainties
of the form factors $F_{i}$ and $G_{i}$ at $q^{2}=0$ for the processes
of $\Xi_{b}\to\Xi_{c}$ and $\Lambda_{b}\to\Lambda_{c}$ are then
given in Table \ref{Tab:F0_central_err}. In this table, central
values of the Borel parameters $(T_{1}^{2},T_{2}^{2})$
are respectively taken as $(25,3)\ {\rm GeV}^{2}$ and $(20,3)\ {\rm GeV}^{2}$ for $\Xi_{b}\to\Xi_{c}$ and $\Lambda_{b}\to\Lambda_{c}$. The three uncertainties are from the Borel region, the threshold parameters $s_{1}^{0}$ and $s_{2}^{0}$ respectively. When determining the uncertainties from the threshold parameters, the suboptimal values in Table \ref{Tab:pole_residues} are used as references.

\begin{table}
\caption{Central values and uncertainties of the form factors $F_{i}$ and
$G_{i}$ at $q^{2}=0$.}
\label{Tab:F0_central_err} %
\begin{tabular}{c|c|c|c|c|c}
\hline
Transition  & $F$  & Central value & Err from $T_{1,2}^{2}$ & Err from $s_{1}^{0}$ & Err from $s_{2}^{0}$\tabularnewline
\hline
 & $F_{1}(0)$  & $-0.125$  & $0.020$ & $0.000$ & $0.010$\tabularnewline
 & $F_{2}(0)$  & $-0.067$  & $0.005$ & $0.004$ & $0.002$\tabularnewline
$\Xi_{b}\to\Xi_{c}$  & $F_{3}(0)$  & $0.701$  & $0.043$ & $0.022$ & $0.032$\tabularnewline
 & $G_{1}(0)$  & $-0.156$  & $0.021$ & $0.002$ & $0.014$\tabularnewline
 & $G_{2}(0)$  & $0.096$  & $0.004$ & $0.005$ & $0.004$\tabularnewline
 & $G_{3}(0)$  & $0.518$  & $0.036$ & $0.023$ & $0.025$\tabularnewline
\hline
 & $F_{1}(0)$  & $-0.101$  & $0.010$ & $0.001$ & $0.016$\tabularnewline
 & $F_{2}(0)$  & $-0.059$  & $0.009$ & $0.004$ & $0.004$\tabularnewline
$\Lambda_{b}\to\Lambda_{c}$  & $F_{3}(0)$  & $0.604$  & $0.098$ & $0.023$ & $0.056$\tabularnewline
 & $G_{1}(0)$  & $-0.124$  & $0.012$ & $0.002$ & $0.019$\tabularnewline
 & $G_{2}(0)$  & $0.080$  & $0.011$ & $0.005$ & $0.007$\tabularnewline
 & $G_{3}(0)$  & $0.456$  & $0.079$ & $0.019$ & $0.036$\tabularnewline
\hline
\end{tabular}
\end{table}

To access the $q^{2}$ dependence of the form factors, we calculate the form factors in a small interval $q^{2}\in [0.0,0.5] \ {\rm GeV}^{2}$, and fit the values with the following simplified $z$-expansion \cite{Detmold:2015aaa}:
\begin{equation}
f(q^{2})=\frac{a+b\ z(q^{2})}{1-q^{2}/m_{{\rm pole}}^{2}},
\end{equation}
where $m_{{\rm pole}}= m_{B_{c}}$,
\begin{equation}
z(q^{2})=\frac{\sqrt{t_{+}-q^{2}}-\sqrt{t_{+}-q_{{\rm max}}^{2}}}{\sqrt{t_{+}-q^{2}}+\sqrt{t_{+}-q_{{\rm max}}^{2}}}
\end{equation}
with $t_{+}=m_{{\rm pole}}^{2}$ and $q_{{\rm max}}^{2}=(M_{1}-M_{2})^{2}$,
$M_{1}=m_{\Xi_{b}}$ ($m_{\Lambda_{b}}$), $M_{2}=m_{\Xi_{c}}$ ($m_{\Lambda_{c}}$). The fitted results of $(a,b)$ for the form factors of $\Xi_{b}\to\Xi_{c}$
and $\Lambda_{b}\to\Lambda_{c}$ are given in Table \ref{Tab:a_b}. In Table \ref{Tab:comparison_ff}, our results are compared with those of the Lattice QCD \cite{Detmold:2015aaa} and those of HQET at the next-to-leading power (NLP) of $1/m_{Q}$ \cite{Manohar:2000dt}. Some comments are in order.
\begin{itemize}

\item For the results of HQET at the NLP in Table \ref{Tab:comparison_ff}, we have used
\begin{equation}
\zeta(1)=1,\quad\bar{\Lambda}_{\Lambda}=0.9\ {\rm GeV},\quad m_{c}=1.4\ {\rm GeV},\quad m_{b}=4.8\ {\rm GeV}.
\end{equation}
The evaluation of nonperturbative constant $\bar{\Lambda}_{\Lambda}$ can be found in \cite{Zhao:2020wbw} .

\item As can be seen in Table \ref{Tab:F0_central_err}
that, about 10-20\% uncertainties are introduced for the form factors
of $\Xi_{b}\to\Xi_{c}$ at $q^{2}=0$, and about 20\% for $\Lambda_{b}\to\Lambda_{c}$.
As a rough estimate, we assume that similar uncertainties are implicit
in our predictions for the form factors at $q^{2}=q_{{\rm max}}^{2}$ in Table \ref{Tab:comparison_ff}.
Considering these uncertainties, one can see that our results are very
close to those of the Lattice QCD and HQET at NLP, especially for $F_{3}$
and $G_{3}$.

\item As stated above, the form factors of $\Lambda_{b}\to\Lambda_{c}$
at the LP of HQET is:
\begin{equation}
F_{1}=F_{2}=G_{1}=G_{2}=0,\quad F_{3}=G_{3}=1,
\end{equation}
Compared with which, about 10-40\% corrections are introduced for
the NLP results. Since the mass of charm quark is not high enough, both HQET and HQET sum rules may receive sizable power corrections.

\end{itemize}
\begin{table}
\caption{The fitted results of $(a,b)$ for the form factors.}
\label{Tab:a_b} %
\begin{tabular}{c|c|c||c|c|c}
\hline
Transition  & $F$  & $(a,b)$ & Transition  & $F$  & $(a,b)$\tabularnewline
\hline
 & $F_{1}$  & $(-0.198,0.891)$  &  & $F_{1}$  & $(-0.195,1.134)$ \tabularnewline
 & $F_{2}$  & $(-0.127,0.735)$  &  & $F_{2}$  & $(-0.121,0.752)$ \tabularnewline
$\Xi_{b}\to\Xi_{c}$  & $F_{3}$  & $(1.053,-4.282)$  & $\Lambda_{b}\to\Lambda_{c}$  & $F_{3}$  & $(1.064,-5.561)$ \tabularnewline
 & $G_{1}$  & $(-0.283,1.546)$  &  & $G_{1}$  & $(-0.248,1.509)$ \tabularnewline
 & $G_{2}$  & $(0.197,-1.225)$  &  & $G_{2}$  & $(0.183,-1.243)$ \tabularnewline
 & $G_{3}$  & $(0.847,-3.961)$  &  & $G_{3}$  & $(0.762,-3.685)$ \tabularnewline
\hline
\end{tabular}
\end{table}

\begin{table}
\caption{Our predictions for the form factors at $q^{2}=0$ and $q^{2}=q_{{\rm max}}^{2}$
are compared with those from the Lattice QCD and the next-to-leading
power of $1/m_{Q}$ in HQET \cite{Manohar:2000dt}. For the latter, only
$F(q_{{\rm max}}^{2})$ are shown, the choices for the parameter values can be found in the text. As can be seen in Table \ref{Tab:F0_central_err},
about $10-20\%$ uncertainties can be introduced in our results for $\Xi_{b}\to\Xi_{c}$
and $\Lambda_{b}\to\Lambda_{c}$.}
\label{Tab:comparison_ff} %
\begin{tabular}{c|c|c|c|c}
\hline
Transition  & $F$  & This work & LQCD \cite{Detmold:2015aaa}  & HQET@NLP \cite{Manohar:2000dt}\tabularnewline
\hline
 & $F_{1}$  & $(-0.125,-0.276)$  & - -  & $-0.321$\tabularnewline
 & $F_{2}$  & $(-0.067,-0.177)$  & - -  & $-0.094$\tabularnewline
$\Xi_{b}\to\Xi_{c}$  & $F_{3}$  & $(0.701,1.464)$  & - -  & $1.415$\tabularnewline
 & $G_{1}$  & $(-0.156,-0.394)$  & - -  & $-0.321$\tabularnewline
 & $G_{2}$  & $(0.096,0.274)$  & - -  & $0.094$\tabularnewline
 & $G_{3}$  & $(0.518,1.178)$  & - -  & $1$\tabularnewline
\hline
 & $F_{1}$  & $(-0.101,-0.271)$  & $(-0.174,-0.419)$  & $-0.321$\tabularnewline
 & $F_{2}$  & $(-0.059,-0.168)$  & $(-0.010,-0.086)$  & $-0.094$\tabularnewline
$\Lambda_{b}\to\Lambda_{c}$  & $F_{3}$  & $(0.604,1.482)$  & $(0.558,1.492)$  & $1.415$\tabularnewline
 & $G_{1}$  & $(-0.124,-0.346)$  & $(-0.210,-0.493)$  & $-0.321$\tabularnewline
 & $G_{2}$  & $(0.080,0.255)$  & $(0.082,0.196)$  & $0.094$\tabularnewline
 & $G_{3}$  & $(0.456,1.061)$  & $(0.388,0.907)$  & $1$\tabularnewline
\hline
\end{tabular}
\end{table}

The predictions for the form factors are then applied to the semi-leptonic
processes, and we arrive at:
\begin{align}
\Gamma & =(3.80\pm0.33)\times10^{-14}\ {\rm GeV},\nonumber \\
{\cal B} & =(9.02\pm0.79)\%,\nonumber \\
\Gamma_{L}/\Gamma_{T} & =1.29\pm0.06
\end{align}
for $\Xi_{b}\to\Xi_{c}e^{-}\bar{\nu}_{e}$, and
\begin{align}
\Gamma & =(2.96\pm0.48)\times10^{-14}\ {\rm GeV},\nonumber \\
{\cal B} & =(6.61\pm1.08)\%,\nonumber \\
\Gamma_{L}/\Gamma_{T} & =1.28\pm0.12
\end{align}
for $\Lambda_{b}\to\Lambda_{c}e^{-}\bar{\nu}_{e}$. The uncertainties
come from those of the form factors. Our predictions for the branching fractions are compared with
those from the Lattice QCD \cite{Detmold:2015aaa} and the experimental data \cite{Tanabashi:2018oca}, as can be seen in Table \ref{Tab:comparison_semi_lep}. It can be seen that our prediction for $\Lambda_{b}\to\Lambda_{c}e^{-}\bar{\nu}_{e}$ is consistent with that of the Lattice QCD and the experiment. In addition, the SU(3) symmetry breaking between $\Xi_{b}\to\Xi_{c}e^{-}\bar{\nu}_{e}$ and $\Lambda_{b}\to\Lambda_{c}e^{-}\bar{\nu}_{e}$ is about 30\%.

\begin{table}
\caption{Our predictions for the semi-leptonic branching fractions (in units of $\%$) are compared with those from the Lattice
QCD \cite{Detmold:2015aaa} and the experimental data \cite{Tanabashi:2018oca}.}
\label{Tab:comparison_semi_lep} %
\begin{tabular}{c|c|c|c}
\hline
Channel  & This work  & Lattice QCD \cite{Detmold:2015aaa}  & Experimental data \cite{Tanabashi:2018oca} \tabularnewline
\hline
$\Xi_{b}\to\Xi_{c}e^{-}\bar{\nu}_{e}$  & $9.02\pm0.79$  & - - & - -\tabularnewline
$\Lambda_{b}\to\Lambda_{c}e^{-}\bar{\nu}_{e}$  & $6.61\pm1.08$  & $5.32\pm0.35$  & $6.2_{-1.3}^{+1.4}$ \tabularnewline
\hline
\end{tabular}
\end{table}

\section{Conclusions}

In this work, the full LO results of the form factors for the processes  $\Xi_{b}\to\Xi_{c}$ and $\Lambda_{b}\to\Lambda_{c}$ are obtained in QCD sum rules. For completeness, we also study the two-point correlation functions to obtain the pole residues of $\Xi_{Q}$ and $\Lambda_{Q}$. Contributions from up to dim-5 operators have been considered. We have also included the leading logarithm approximation.  For the two-point correlation function, since the perturbative contribution dominates and a stable Borel window for the pole residue can be found, higher accuracy is achieved both for the pole residue and mass. However, although the perturbative contribution also dominates for the three-point correlation function, a stable Borel region can hardly be found. Somewhat artificial criteria has to be adopted to select the relatively optimal region, and about 20\% uncertainties are introduced for the form factors of $\Xi_{b}\to\Xi_{c}$ and $\Lambda_{b}\to\Lambda_{c}$. Our results of the form factors are consistent with those of the Lattice QCD within errors. It is worth noting that, starting from our full LO results, one can arrive at the results of HQET sum rules, when $m_{b,c}$ are taken to be infinity \cite{Shuryak:1981fza,Zhao:2020wbw}. Similar arguments have been performed in \cite{MarquesdeCarvalho:1999bqs}.

In \cite{Shi:2019hbf}, we derived the form factors of doubly heavy baryons to singly heavy baryons using QCD sum rules for the first time, but so far our results are hard to be tested for the lack of experimental data. For the singly heavy baryon decays more data are accumulated which can help to check the theoretical predictions. In this work reliable results are obtained for the form factors of $\Xi_{b}\to\Xi_{c}$ and $\Lambda_{b}\to\Lambda_{c}$ so that the validity of the three-point sum rules is checked. A potential application is to calculate the matrix elements for the lifetimes of weakly-decay heavy baryons. Our forthcoming works will focus on this problem.

\section*{Acknowledgements}

The authors are grateful to Profs. Wei Wang and Zhi-Gang Wang for valuable discussions. This work is supported in part by National Natural Science Foundation of China under Grants No. 11765012, 11947414.

\end{document}